\documentclass[%
 reprint,
 amsmath,amssymb,
 aps,
prl,
]{revtex4-2}
\usepackage{graphicx}
\usepackage{dcolumn}
\usepackage{bm}
\usepackage{svg}

\usepackage{xcolor}
\begin{document}

\title{Robust Critical Connectivity Threshold in Ranked Percolation of Granular Packings}

\author{Vasco C. Braz}
\email{vcbraz@fc.ul.pt}
\affiliation{Centro de Física Teórica e Computacional, Faculdade de Ciências da Universidade de Lisboa, 1749-016 Lisboa, Portugal}
\affiliation{Departamento de Física, Faculdade de Ciências da Universidade de Lisboa, 1749-016 Lisboa, Portugal}
\author{N. A. M. Araújo}
\email{nmaraujo@fc.ul.pt}
\affiliation{Centro de Física Teórica e Computacional, Faculdade de Ciências da Universidade de Lisboa, 1749-016 Lisboa, Portugal}
\affiliation{Departamento de Física, Faculdade de Ciências da Universidade de Lisboa, 1749-016 Lisboa, Portugal}

\date{\today}

\begin{abstract}
The formation of sintering bridges in amorphous powders affects both flow behavior and perceived material quality. When sintering is driven by surface tension, bridges emerge sequentially, favoring contacts between smaller particles first. Predicting the connectivity percolation threshold is key to understanding and controlling the onset of sintering. We investigate ranked percolation in granular packings, where particles connect based on contact number. While the percolation threshold defined by the fraction of connected particles is non-universal and highly sensitive to particle size dispersion, we find that the critical number of sintered bridges per particle provides a robust estimator across different size distributions. Through numerical simulations and a mean-field approach, we link this robustness to the spatial distribution of contacts. Our results have broader implications for understanding the resilience of spatially embedded networks under targeted attacks.
\end{abstract}

\maketitle

Granular materials are ubiquitous in daily life, appearing in systems ranging from rice grains and coffee powder to sand dunes, chemical compounds, and fertilizers~\cite{Jaeger1996,Duran2001,Nagel2017}. Depending on the properties of individual grains and external constraints, whether natural or engineered, these systems typically form disordered packings. The mechanical behavior of such packings is intimately tied to their microstructure. Geometric features, such as particle size, shape, and spatial distribution, define the structure of the contact network, which in turn governs the rheological response of the material~\cite{Torquato2002,Smart2008,Papadopoulos2018,Behringer2019}. Mechanical stability and stress propagation are sustained by force chains whose topology depends on particle size distribution, spatial arrangement, and interparticle interactions~\cite{Liu1995,Mueth1998,Walker2012,Baule2018,Papadopoulos2016,Krishnaraj2020,Mandal2022}.

A central factor influencing the mechanical properties of granular materials is the number of contacts per particle. In non-cohesive granular systems, the stiffness of a packing can often be correlated with the average coordination number~\cite{Alexander1998,Wyart2005}. These interactions are typically mediated by Coulomb friction and repulsive Hertzian contacts. However, in many practical situations, cohesive forces become dominant~\cite{Forsyth01,Gans24,Fujio2024,Shrivastava25,Ness25,Sharma2025}. For instance, in wet granular media, capillary bridges form resulting in attractive forces between neighboring grains~\cite{Jaeger1996,Hornbaker1997,Selmani24}. Cohesive interparticle interactions also arise in amorphous molecular powders such as instant coffee or table sugar. Temperature- or moisture-induced plasticization in these systems can promote the formation of sintered ``bridges'' at particle contacts. In both amorphous and crystalline amorphous molecular powders, such cohesive bridges alter the mechanical behavior of the packing and can eventually lead to aggregation into multiple ``lumps'' or even a fully connected solid structure, commonly referred to as ``caking''~\cite{Palzer2005,Hartmann2011,Zafar2017,Chen2019,Zaccone2022,Middleton2022,Xu2025}.

\begin{figure*}[t]
\centering
\includegraphics[width=1\textwidth]{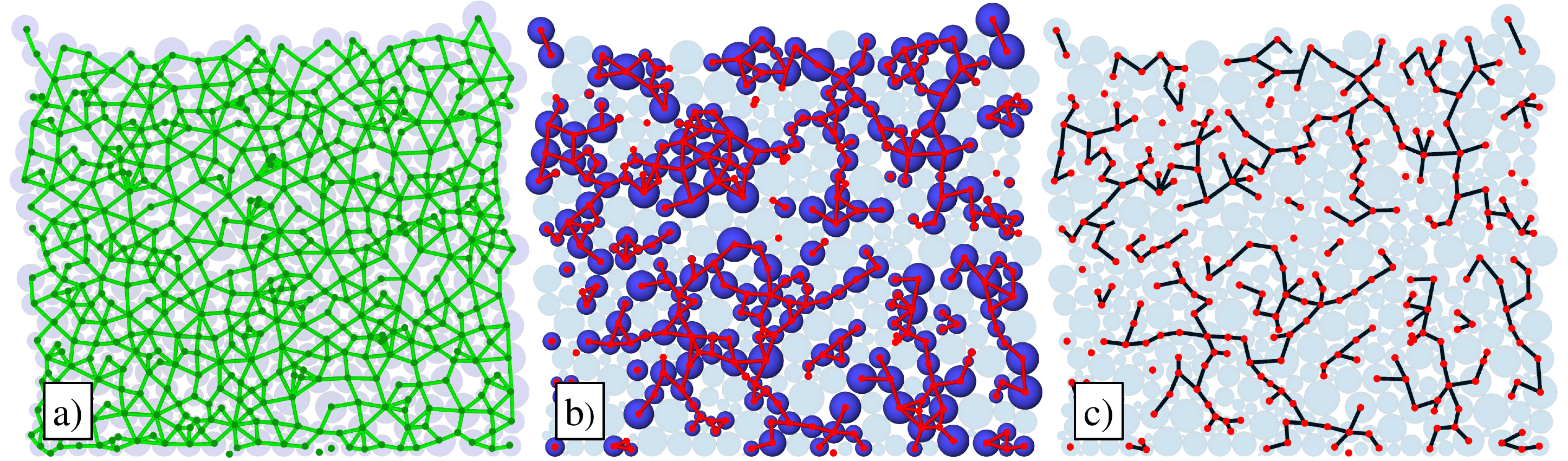}
\caption{\textbf{Contact, cohesive, and minimum spanning forest networks in granular packings.} (a) Contact network of a bed of $500$ spherical particles (light blue spheres), where each node (green dot) is placed at the center of a particle, and edges (green lines) connect nodes representing pairs of touching particles.
(b) Cohesive network formed after 60\% of nodes are occupied in order of their degree. This is a subgraph of the contact network shown in (a), containing only the particles connected by cohesive (sintered) bridges.
(c) Minimum Spanning Forest: a subgraph of the cohesive network where only the minimum spanning trees of each cluster are retained. Edge weights are assigned based on the highest degree among the connected nodes.
\label{fgr:contact_network}}
\end{figure*}

To describe the topology of the caking process, we consider two distinct graphs. The first is the \textit{contact network}, constructed by representing each particle as a node and each physical contact between particles as an edge (Fig.\ref{fgr:contact_network}a). The second is the \textit{cohesive network}, a subgraph of the contact network that includes only particles connected by cohesive (sintered) bridges and the corresponding edges (Fig.\ref{fgr:contact_network}b). The evolution of the cohesive network is not random; it depends on the physical mechanisms and the history of bridge formation. Recent studies have shown that the caking of amorphous powders can be mapped onto a percolation process in which node occupation follows a well-defined order, governed by the underlying sintering dynamics~\cite{Braz2022,Simoes2022}. In this framework, each particle is either occupied (cohesively connected) or absent from the cohesive network. As the fraction of occupied nodes $p$ increases, the network transitions from a fragmented state to one in which a large, system-spanning cluster appears. This transition occurs at a critical occupation threshold $p_c$, marking the onset of connectivity percolation~\cite{Artime2024}.

The order in which nodes are occupied is dictated by the underlying physical mechanism and directly influences the critical occupation fraction at which percolation emerges~\cite{Araujo2014b}. When particle aggregation is triggered by a humidity shock, caking can be mapped onto a ranked percolation process where nodes are occupied in order of increasing particle size~\cite{Braz2022}. In contrast, under a temperature shock, the ordering is altered, and smaller grains may even be excluded from the sintering process~\cite{Simoes2022}. In densely packed granular materials, all contacts occur at particle surfaces, and larger particles typically have more contacts due to their greater surface area. This results in a strong correlation between particle size and contact number~\cite{MuirWood2008,Azema2017}. In this work, we investigate how size dispersion in granular beds influences the percolation threshold when nodes are occupied in order of increasing connectivity, from the least connected (often the smallest grains) to the most connected (often the largest). We demonstrate that, despite the non-universality of the percolation threshold itself, the average number of connections per occupied node at the critical point serves as a robust and consistent indicator of the onset of connectivity percolation.

\begin{figure}[t]
\centering
\includegraphics[width=1\columnwidth]{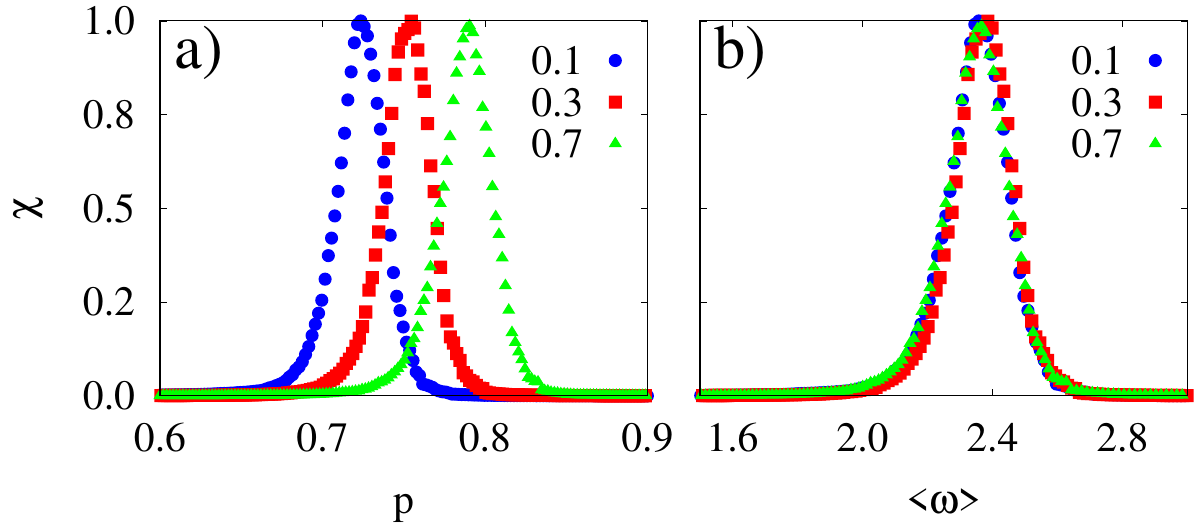}%
\caption{\textbf{Fluctuations of the largest cluster size.} Evolution of the fluctuations of the largest cluster size $\phi$, normalized by its maximum value, in 2D random packs for \textit{Ascending Degree Ranked Percolation} for size distributions with a standard deviation $\sigma$ of $0.1$ (blue), $0.3$ (red), and $0.7$ (green). The fluctuations are shown as a function of (a) the occupation fraction $p$ and (b) the number of connections per occupied node $\langle \omega \rangle$.}
\label{fgr:fluctuations}
\end{figure}

We consider granular beds resulting from the sedimentation of particles with radii that follow a normal size distribution, $p(r)$, with mean $\langle r \rangle = 1$ and standard deviation $\sigma$ varying between zero (monodisperse) and one. We generated 500 compact packings, each containing 5000 particles, for every value of $\sigma$. The topology of the resulting contact network naturally depends on the heterogeneity of the original particle bed. For low size dispersion (low values of $\sigma$), most nodes have a degree (i.e., number of contacts) of $k = 2D$, where $D$ is the spatial dimension. For larger dispersions, smaller particles tend to contact larger ones in order to maximize the packing fraction. As a result, the degree distribution of the contact network exhibits increased kurtosis, and the average node degree grows monotonically with particle size (see Figs.~S1 and S2 in the Supplemental Material~\cite{SM}).

We rank each node according to its degree. To evaluate the \textit{Degree-Ranked Percolation} threshold for each sample, we consider varying levels of node occupation. Starting from an empty network (with all nodes and their corresponding edges removed from the original contact network), we sequentially introduce nodes based on their degree rank and measure the properties of the evolving network. In this process, all nodes with degree $k = 0$ are first occupied at random, followed by those with $k = 1$, and so on. At each step, we associate with the original contact network of $N$ nodes a sequence of $N$ \textit{cohesive networks} (Fig.~\ref{fgr:contact_network}b), each representing a subgraph of the original contact network. 

We analyze the percolation transition by measuring the fluctuations in the size of the largest cluster, $\phi$, defined as
$\chi=\frac{\langle\phi^2\rangle -\langle\phi\rangle^2}{[\langle\phi^2\rangle -\langle\phi\rangle^2]_{max}}$, which is normalized by the maximum value observed across all values of the fraction of nodes occupied $p$. Figure~\ref{fgr:fluctuations}a shows the dependence of $\chi$ on $p$. The percolation threshold, estimated from the position of the peak in $\chi$, increases with $\sigma$. When nodes are occupied in order of increasing degree, the resulting cohesive network has lower connectivity at a given $p$ compared to a network formed by random node occupation with the same number of occupied nodes.

\begin{figure}[t]
	\centering
\includegraphics[width=1.0\columnwidth]{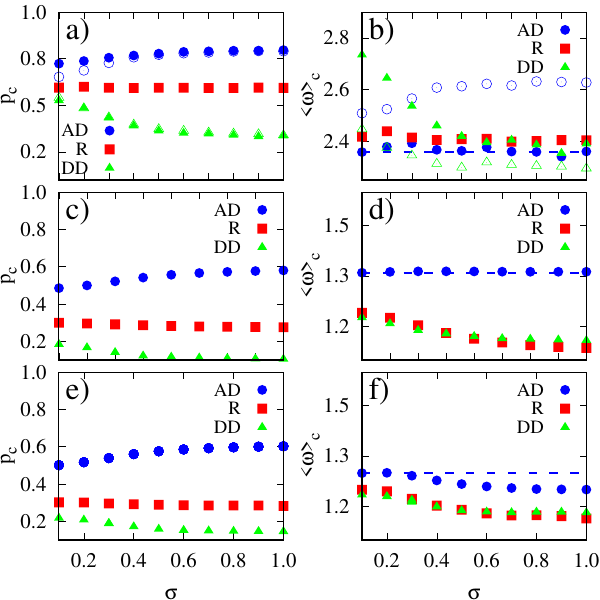}
\caption{
\textbf{Onset of percolation for random particle packs, configuration model, and degree-correlated synthetic networks.}
Dependence of the percolation threshold and the critical number of connections per node on the size dispersion~$\sigma$ for random particle packs (a, b), the configuration model (c, d), and degree-correlated synthetic network (e, f). Different symbols indicate different percolation models: blue filled circles (Ascending Degree Ranked Percolation), red filled squares (Random Percolation), and green filled triangles (Descending Degree Ranked Percolation). In (a) and (b), open blue circles and open green triangles represent Ascending and Descending Size Ranked Percolation, respectively.
}
	\label{fgr:critical}
\end{figure}

To compare the state of different cohesive networks derived from contact networks with varying size dispersions but the same average connectivity, we plotted the fluctuations as a function of the average number of connections per occupied node $\langle \omega \rangle$ and observed that all curves collapse onto a single peak around $\langle \omega \rangle_c \approx 2.36$. This suggests that $\langle \omega \rangle_c$ is a robust threshold for the degree-ranked percolation process under consideration (Fig.~\ref{fgr:fluctuations}b). In Fig.~\ref{fgr:critical}, we plot the percolation threshold (Fig.~\ref{fgr:critical}a) and the corresponding critical number of connections per node (Fig.~\ref{fgr:critical}b) for different size dispersions. It is evident that, while the percolation threshold increases with $\sigma$, the critical number of connections per node remains approximately constant when occupation is performed in ascending order of degree (AD, blue dots). However, this invariance does not hold for descending degree-ranked percolation (DD, green triangles), where nodes are occupied from the most to the least connected. In that case, increasing the size dispersion results in a larger number of high-degree nodes, leading to a decrease in both the percolation threshold and the critical number of connections per node (Fig.~\ref{fgr:critical}a). Additionally, the figures show that size-ranked percolation (open symbols) qualitatively mirrors the behavior of degree-ranked percolation. For low size dispersions, where particles have nearly uniform radii, both thresholds approach those observed in random percolation. As size (and degree) dispersion increases, the correlation between particle size and connectivity becomes more pronounced. Consequently, the qualitative features of size-ranked percolation can be understood in terms of the corresponding trends observed in degree-ranked percolation, in both ascending and descending order.

To evaluate the applicability and robustness of this result, we investigated the role of spatial and degree-degree correlations present in the original contact networks. To this end, we employed the configuration model as an analytical framework to study networks that share the same degree distribution as the contact graphs for each value of $\sigma$. This approach assumes a locally tree-like structure (i.e., it neglects short loops). While preserving the original degree distributions, we randomized the connections by assigning equal probability for each node to connect to any other, thereby removing the spatial embedding and degree-degree correlations present in the original networks. This procedure allows us to isolate and assess the specific contribution of the degree distribution to the emergence of a giant connected component.

To identify the onset of percolation, as detailed in the Supplemental Material (Section~2.B~\cite{SM}), we derive the degree distribution $q_w$ of the evolving cohesive network and apply the criterion of Molloy and Reed~\cite{Molloy1995} to obtain an analytical expression for the critical average number of connections per node:
\begin{equation}
\langle \omega \rangle_c = \frac{\left[ \left( \langle k \rangle - \alpha_1(k^*) \frac{k^*}{k^{*2} - k^*} \right) - k^* \alpha_2(k^*) \right]^2}{2k \left( k - \alpha_1(k^*) \right) \frac{k^*}{k^{*2} - k^*}} \, ,
\label{Eq:wc_ranked}
\end{equation}
where $k^* = F^{-1}(p)$ is the degree of the highest-degree occupied node, $F$ is the cumulative degree distribution, $\alpha_1(k^*) = \sum_{k=0}^{k^*} p_k (k^2 - k + k^{*2} - k^*) $, and $\alpha_2(k^*) = \sum_{k=0}^{k^*} p_k (k - k^*)$.

\begin{figure}[t]
	\centering
\includegraphics[width=1\columnwidth]{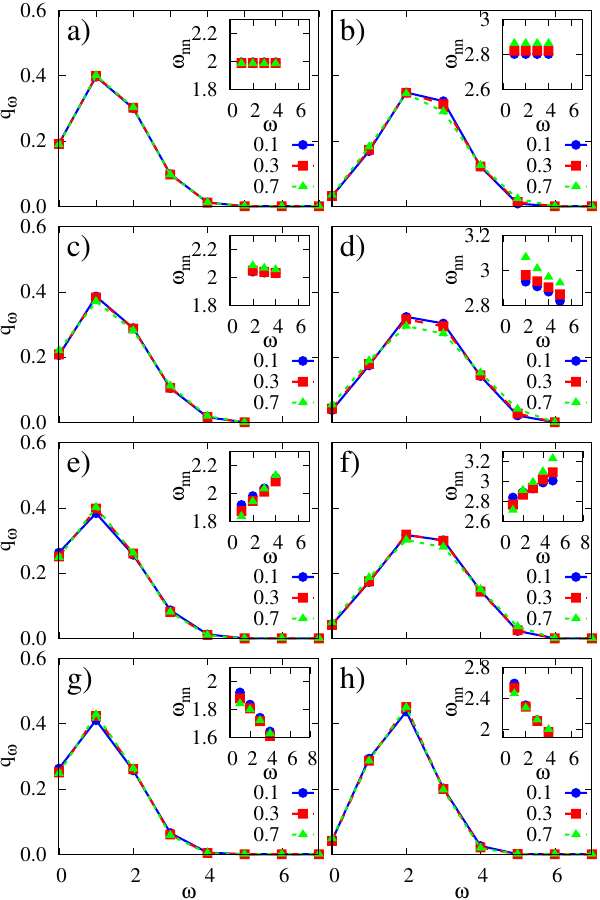}
\caption{\textbf{Degree distributions and correlations for different networks.} Degree distributions $q_\omega$ and degree correlation function $\omega_{nn}$ at $\langle\omega\rangle = 1.33$ (left column) and $\langle\omega\rangle = 2.4$ (right column) for:  
(a) and (b): uncorrelated networks;  
(c) and (d): degree-correlated networks;   
(e) and (f): dandom pack;  
(g) and (h): the set of minimum spanning trees in the random pack. 
 Different symbols indicate different percolation models: blue filled circles (Ascending Degree Ranked Percolation), red filled squares (Random Percolation), and green filled triangles (Descending Degree Ranked Percolation).
}
	\label{fgr:Mean_field}
\end{figure}

Figure~\ref{fgr:critical}d shows that the critical number of connections per node in the uncorrelated networks remains approximately constant for the considered ranked percolation process, but varies significantly when nodes are occupied in reverse order or randomly. Surprisingly, by explicitly calculating the degree distributions $q_w(\omega)$ for values of $\omega \approx \omega_c$, we find that the relationship between the first and second moments of the distribution during ranked occupation across different contact networks arises from the collapse of the corresponding distributions $q_w(\langle \omega \rangle)$ (Fig.~\ref{fgr:Mean_field}a). This indicates that cohesive networks with $\langle \omega \rangle$ near the critical value $\langle \omega \rangle_c$ are topologically equivalent, regardless of the degree dispersion of the original contact network. In Fig~\ref{fgr:critical}a shows that the percolation threshold in the uncorrelated networks as the dispersion increases follows the same qualitative behavior found in the random packs.

The collapse of the degree distributions $q_w(\langle \omega \rangle)$ near the critical point can be attributed to a compensation effect between structural and statistical properties of the underlying networks. In highly dispersed contact networks ($\sigma > 0.1$), achieving the same average connectivity as in low-dispersion systems ( $\sigma \approx 0.1$) requires more occupied nodes and edges in the cohesive network. This increases the likelihood that an edge from an occupied node connects to another occupied one. Nodes with degrees near the mean $\langle k \rangle$ tend to retain more of their original connectivity, while the fraction of poorly connected nodes decreases. However, although occupied nodes are more likely to form connections, they generally have fewer available edges. These two effects nearly cancel out around $\langle \omega \rangle_c$, resulting in the observed collapse (see Fig.~S3 in the Supplemental Material~\cite{SM}).

To test whether this robustness persists in correlated networks, we extended our analysis to account for degree-degree correlations present in the original contact graphs. This generalization involves computing the joint degree-degree probability matrix $e_{kk'}$, which preserves not only the degree distribution but also the degrees of neighboring nodes, while still neglecting spatial correlations. From this matrix, we derive both the degree distribution and the joint degree-degree correlation matrix $c_{ij}$ of the cohesive network at a given $\omega$, following the method introduced in~\cite{Peruani2012}. We then construct the corresponding \textit{branching matrix} $ B_{kk'}$ and apply a generalized Molloy–Reed criterion, computing the percolation threshold from the largest eigenvalue of this matrix~\cite{Dorogovtsev2022} (see Supplemental Material~\cite{SM}).

As the degree dispersion $\sigma$ increases, degree correlations also become more pronounced, with highly connected nodes more likely to connect to poorly connected ones (disassortativity).  While the critical percolation threshold $p_c$ exhibits qualitatively similar behavior to that of uncorrelated networks (Fig.\ref{fgr:critical}e), Fig.\ref{fgr:critical}f reveals that as $ \sigma$ increases, the critical number of connections per node decreases. This indicates that degree-degree correlations weaken the compensation effect described earlier. Consequently, at the percolation threshold, the networks no longer exhibit the same topological features.

The fact that $ \omega_c $ is independent of degree dispersion in both uncorrelated networks and random packs, but not in degree-correlated networks, indicates that the specific spatial configurations of the original contact network impose higher-order correlations that effectively recover the behavior of uncorrelated high-dimensional networks. This raises two key questions: (1) What are the main topological features required to observe this phenomenon? (2) How do the topological properties of cohesive networks derived from granular packings differ from those of degree-correlated high-dimensional networks?

In Fig.~\ref{fgr:Mean_field}f, we show that, as in the case of correlated infinite-dimensional networks, the random packs at the percolation threshold exhibit distinct topological characteristics, such as different $q_{\omega}$ and $\omega_{\mathrm{nn}}$ for different values of $\sigma$. In particular, while degree correlations at $\langle \omega_c \rangle$ vary with $\sigma$, they are no longer disassortative, as observed in the original contact networks, but become slightly assortative. To our knowledge, this effect has not previously been reported for spatially embedded networks and warrants further investigation. Although a similar trend was reported in~\cite{Peruani2012} and attributed to homogenization due to the removal of high-degree nodes, our results suggest a different mechanism. Here, the observed assortative correlations appear to result from node removal in highly clustered networks, a natural consequence of spatial embedding.

This behavior is consistently observed across all three percolation models. Furthermore, we found the same effect for percolation on a hexagonal lattice, which has a high clustering coefficient, but not on a square lattice, where the clustering coefficient is zero (see Supplemental Material~\cite{SM}). In highly clustered networks, the presence of triangles plays a crucial role. Within a triangle, all three nodes are mutually connected. When one node is removed, the remaining two (which are also neighbors) each experience a degree reduction of one, introducing a positive correlation between their degrees. In contrast, edge removal does not generate this effect, as it decreases the degrees of two nodes that are not necessarily connected. This suggests that the observed assortative correlations are a direct consequence of the interplay between high clustering and node removal.

The most significant difference between the non-spatial analytical models and the random packs lies in the presence of short loops in the latter, which introduce redundant connections. To explore the structural backbone responsible for connectivity, we focus on the minimum spanning forest (MSF), defined as the collection of minimum spanning trees extracted from each connected component of the cohesive network, where the weight of each edge is given by the degree of the highest-degree node it connects. As a result, the MSF retains only the minimal set of edges required to ensure global connectivity, prioritizing the earliest sintered contacts.

Figure~\ref{fgr:Mean_field}h shows that, unlike the full contact network, the degree distribution and degree correlation function for nodes with degree greater than two (i.e., those contributing to network branching) in the MSF are remarkably similar across different values of size dispersion. Despite substantial variations in the original contact networks, the set of critical spanning trees remains essentially invariant with respect to degree dispersion. The same behavior is observed in 3D random packs across a wide range of dispersions (up to 20\%), as shown in the Supplemental Material~\cite{SM}.
    
\textit{Conclusions:} We have shown that although the percolation threshold in Degree-Ranked Percolation is non-universal and highly sensitive to particle size dispersion, the critical average number of connections per particle is remarkably robust across a wide range of dispersions. This threshold is not exclusive to granular contact networks; it also emerges in uncorrelated, spatially unconstrained networks with the same degree distributions. The origin of this robustness lies in a structural equivalence: the subgraphs formed by the early occupation of low-degree nodes (or equivalently, by the removal of hubs) in both random granular packings and configuration models exhibit similar topological features. These similarities are encoded in the Minimum Spanning Forest of the cohesive network, which remains essentially invariant across size dispersions.  Furthermore, we show that the Size-Ranked Percolation threshold for various particle beds, relevant for predicting the caking time of amorphous powders, closely follows the Degree-Ranked Percolation threshold at high size dispersion, while it converges toward the random percolation threshold at low dispersion.

Degree-Ranked Percolation naturally connects to the broader framework of network resilience under targeted attacks~\cite{Newman2010,Dorogovtsev2022}. Our findings raise fundamental questions about the behavior of spatially embedded networks, which often exhibit narrow, peaked degree distributions~\cite{Barthelemy2011}. Do analogous connectivity thresholds arise in systems such as road networks, power grids, or biological tissues? Addressing these questions is essential for the advancement of applications in fine-powder technologies and to deepen our theoretical understanding of universality, robustness, and failure in complex networked systems.

We acknowledge financial support from the Portuguese Foundation for Science and Technology (FCT) under the contracts: UIDB/00618/2020 (DOI:10.54499/UIDB/00618/2020), UIDP/00618/2020 (DOI:10.54499/UIDP/00618/2020), and 2021.05649.BD.

\appendix

\bibliography{main}

\begin{thebibliography}{43}%
\makeatletter
\providecommand \@ifxundefined [1]{%
 \@ifx{#1\undefined}
}%
\providecommand \@ifnum [1]{%
 \ifnum #1\expandafter \@firstoftwo
 \else \expandafter \@secondoftwo
 \fi
}%
\providecommand \@ifx [1]{%
 \ifx #1\expandafter \@firstoftwo
 \else \expandafter \@secondoftwo
 \fi
}%
\providecommand \natexlab [1]{#1}%
\providecommand \enquote  [1]{``#1''}%
\providecommand \bibnamefont  [1]{#1}%
\providecommand \bibfnamefont [1]{#1}%
\providecommand \citenamefont [1]{#1}%
\providecommand \href@noop [0]{\@secondoftwo}%
\providecommand \href [0]{\begingroup \@sanitize@url \@href}%
\providecommand \@href[1]{\@@startlink{#1}\@@href}%
\providecommand \@@href[1]{\endgroup#1\@@endlink}%
\providecommand \@sanitize@url [0]{\catcode `\\12\catcode `\$12\catcode
  `\&12\catcode `\#12\catcode `\^12\catcode `\_12\catcode `\%12\relax}%
\providecommand \@@startlink[1]{}%
\providecommand \@@endlink[0]{}%
\providecommand \url  [0]{\begingroup\@sanitize@url \@url }%
\providecommand \@url [1]{\endgroup\@href {#1}{\urlprefix }}%
\providecommand \urlprefix  [0]{URL }%
\providecommand \Eprint [0]{\href }%
\providecommand \doibase [0]{https://doi.org/}%
\providecommand \selectlanguage [0]{\@gobble}%
\providecommand \bibinfo  [0]{\@secondoftwo}%
\providecommand \bibfield  [0]{\@secondoftwo}%
\providecommand \translation [1]{[#1]}%
\providecommand \BibitemOpen [0]{}%
\providecommand \bibitemStop [0]{}%
\providecommand \bibitemNoStop [0]{.\EOS\space}%
\providecommand \EOS [0]{\spacefactor3000\relax}%
\providecommand \BibitemShut  [1]{\csname bibitem#1\endcsname}%
\let\auto@bib@innerbib\@empty
\bibitem [{\citenamefont {Jaeger}\ \emph {et~al.}(1996)\citenamefont {Jaeger},
  \citenamefont {Nagel},\ and\ \citenamefont {Behringer}}]{Jaeger1996}%
  \BibitemOpen
  \bibfield  {author} {\bibinfo {author} {\bibfnamefont {H.~M.}\ \bibnamefont
  {Jaeger}}, \bibinfo {author} {\bibfnamefont {S.~R.}\ \bibnamefont {Nagel}},\
  and\ \bibinfo {author} {\bibfnamefont {R.~P.}\ \bibnamefont {Behringer}},\
  }\bibfield  {title} {\bibinfo {title} {Granular solids, liquids, and gases},\
  }\href {https://doi.org/10.1103/RevModPhys.68.1259} {\bibfield  {journal}
  {\bibinfo  {journal} {Rev. Mod. Phys.}\ }\textbf {\bibinfo {volume} {68}},\
  \bibinfo {pages} {1259} (\bibinfo {year} {1996})}\BibitemShut {NoStop}%
\bibitem [{\citenamefont {Duran}\ and\ \citenamefont
  {Behringer}(2001)}]{Duran2001}%
  \BibitemOpen
  \bibfield  {author} {\bibinfo {author} {\bibfnamefont {J.}~\bibnamefont
  {Duran}}\ and\ \bibinfo {author} {\bibfnamefont {R.}~\bibnamefont
  {Behringer}},\ }\bibfield  {title} {\bibinfo {title} {Sands, powders, and
  grains: An introduction to the physics of granular materials},\ }\href
  {https://doi.org/10.1063/1.1383168} {\bibfield  {journal} {\bibinfo
  {journal} {Phys. Today}\ }\textbf {\bibinfo {volume} {54}},\ \bibinfo {pages}
  {63} (\bibinfo {year} {2001})}\BibitemShut {NoStop}%
\bibitem [{\citenamefont {Nagel}(2017)}]{Nagel2017}%
  \BibitemOpen
  \bibfield  {author} {\bibinfo {author} {\bibfnamefont {S.~R.}\ \bibnamefont
  {Nagel}},\ }\bibfield  {title} {\bibinfo {title} {Experimental soft-matter
  science},\ }\href {https://doi.org/10.1103/RevModPhys.89.025002} {\bibfield
  {journal} {\bibinfo  {journal} {Rev. Mod. Phys.}\ }\textbf {\bibinfo {volume}
  {89}},\ \bibinfo {pages} {025002} (\bibinfo {year} {2017})}\BibitemShut
  {NoStop}%
\bibitem [{\citenamefont {Torquato}(2002)}]{Torquato2002}%
  \BibitemOpen
  \bibfield  {author} {\bibinfo {author} {\bibfnamefont {S.}~\bibnamefont
  {Torquato}},\ }\href {https://doi.org/10.1007/978-1-4757-6355-3} {\emph
  {\bibinfo {title} {Random Heterogeneous Materials: Microstructure and
  Macroscopic Properties}}}\ (\bibinfo  {publisher} {Springer},\ \bibinfo
  {address} {New York},\ \bibinfo {year} {2002})\BibitemShut {NoStop}%
\bibitem [{\citenamefont {Smart}\ and\ \citenamefont
  {Ottino}(2008)}]{Smart2008}%
  \BibitemOpen
  \bibfield  {author} {\bibinfo {author} {\bibfnamefont {A.}~\bibnamefont
  {Smart}}\ and\ \bibinfo {author} {\bibfnamefont {J.~M.}\ \bibnamefont
  {Ottino}},\ }\bibfield  {title} {\bibinfo {title} {Granular matter and
  networks: Three related examples},\ }\href {https://doi.org/10.1039/B802672F}
  {\bibfield  {journal} {\bibinfo  {journal} {Soft Matter}\ }\textbf {\bibinfo
  {volume} {4}},\ \bibinfo {pages} {2125} (\bibinfo {year} {2008})}\BibitemShut
  {NoStop}%
\bibitem [{\citenamefont {Papadopoulos}\ \emph {et~al.}(2018)\citenamefont
  {Papadopoulos}, \citenamefont {Porter}, \citenamefont {Daniels},\ and\
  \citenamefont {Bassett}}]{Papadopoulos2018}%
  \BibitemOpen
  \bibfield  {author} {\bibinfo {author} {\bibfnamefont {L.}~\bibnamefont
  {Papadopoulos}}, \bibinfo {author} {\bibfnamefont {M.~A.}\ \bibnamefont
  {Porter}}, \bibinfo {author} {\bibfnamefont {K.~E.}\ \bibnamefont
  {Daniels}},\ and\ \bibinfo {author} {\bibfnamefont {D.~S.}\ \bibnamefont
  {Bassett}},\ }\bibfield  {title} {\bibinfo {title} {Network analysis of
  particles and grains},\ }\href {https://doi.org/10.1093/comnet/cny005}
  {\bibfield  {journal} {\bibinfo  {journal} {J. Complex Netw.}\ }\textbf
  {\bibinfo {volume} {6}},\ \bibinfo {pages} {485} (\bibinfo {year}
  {2018})}\BibitemShut {NoStop}%
\bibitem [{\citenamefont {Behringer}\ and\ \citenamefont
  {Chakraborty}(2019)}]{Behringer2019}%
  \BibitemOpen
  \bibfield  {author} {\bibinfo {author} {\bibfnamefont {R.~P.}\ \bibnamefont
  {Behringer}}\ and\ \bibinfo {author} {\bibfnamefont {B.}~\bibnamefont
  {Chakraborty}},\ }\bibfield  {title} {\bibinfo {title} {The physics of
  jamming for granular materials: A review},\ }\href
  {https://doi.org/10.1088/1361-6633/aadc3c} {\bibfield  {journal} {\bibinfo
  {journal} {Rep. Prog. Phys.}\ }\textbf {\bibinfo {volume} {82}},\ \bibinfo
  {pages} {012601} (\bibinfo {year} {2019})}\BibitemShut {NoStop}%
\bibitem [{\citenamefont {Liu}\ \emph {et~al.}(1995)\citenamefont {Liu},
  \citenamefont {Nagel}, \citenamefont {Schecter}, \citenamefont {Coppersmith},
  \citenamefont {Majumdar}, \citenamefont {Narayan},\ and\ \citenamefont
  {Witten}}]{Liu1995}%
  \BibitemOpen
  \bibfield  {author} {\bibinfo {author} {\bibfnamefont {C.}~\bibnamefont
  {Liu}}, \bibinfo {author} {\bibfnamefont {S.}~\bibnamefont {Nagel}}, \bibinfo
  {author} {\bibfnamefont {D.}~\bibnamefont {Schecter}}, \bibinfo {author}
  {\bibfnamefont {S.}~\bibnamefont {Coppersmith}}, \bibinfo {author}
  {\bibfnamefont {S.}~\bibnamefont {Majumdar}}, \bibinfo {author}
  {\bibfnamefont {O.}~\bibnamefont {Narayan}},\ and\ \bibinfo {author}
  {\bibfnamefont {T.}~\bibnamefont {Witten}},\ }\bibfield  {title} {\bibinfo
  {title} {Force fluctuations in bead packs},\ }\href
  {https://doi.org/10.1126/science.269.5223.513} {\bibfield  {journal}
  {\bibinfo  {journal} {Science}\ }\textbf {\bibinfo {volume} {269}},\ \bibinfo
  {pages} {513} (\bibinfo {year} {1995})}\BibitemShut {NoStop}%
\bibitem [{\citenamefont {Mueth}\ \emph {et~al.}(1998)\citenamefont {Mueth},
  \citenamefont {Jaeger},\ and\ \citenamefont {Nagel}}]{Mueth1998}%
  \BibitemOpen
  \bibfield  {author} {\bibinfo {author} {\bibfnamefont {D.~M.}\ \bibnamefont
  {Mueth}}, \bibinfo {author} {\bibfnamefont {H.~M.}\ \bibnamefont {Jaeger}},\
  and\ \bibinfo {author} {\bibfnamefont {S.~R.}\ \bibnamefont {Nagel}},\
  }\bibfield  {title} {\bibinfo {title} {Force distribution in a granular
  medium},\ }\href {https://doi.org/10.1103/PhysRevE.57.3164} {\bibfield
  {journal} {\bibinfo  {journal} {Phys. Rev. E}\ }\textbf {\bibinfo {volume}
  {57}},\ \bibinfo {pages} {3164} (\bibinfo {year} {1998})}\BibitemShut
  {NoStop}%
\bibitem [{\citenamefont {Walker}\ and\ \citenamefont
  {Tordesillas}(2012)}]{Walker2012}%
  \BibitemOpen
  \bibfield  {author} {\bibinfo {author} {\bibfnamefont {D.~M.}\ \bibnamefont
  {Walker}}\ and\ \bibinfo {author} {\bibfnamefont {A.}~\bibnamefont
  {Tordesillas}},\ }\bibfield  {title} {\bibinfo {title} {Taxonomy of granular
  rheology from grain property networks},\ }\href
  {https://doi.org/10.1103/PhysRevE.85.011304} {\bibfield  {journal} {\bibinfo
  {journal} {Phys. Rev. E}\ }\textbf {\bibinfo {volume} {85}},\ \bibinfo
  {pages} {011304} (\bibinfo {year} {2012})}\BibitemShut {NoStop}%
\bibitem [{\citenamefont {Baule}\ \emph {et~al.}(2018)\citenamefont {Baule},
  \citenamefont {Morone}, \citenamefont {Herrmann},\ and\ \citenamefont
  {Makse}}]{Baule2018}%
  \BibitemOpen
  \bibfield  {author} {\bibinfo {author} {\bibfnamefont {A.}~\bibnamefont
  {Baule}}, \bibinfo {author} {\bibfnamefont {F.}~\bibnamefont {Morone}},
  \bibinfo {author} {\bibfnamefont {H.~J.}\ \bibnamefont {Herrmann}},\ and\
  \bibinfo {author} {\bibfnamefont {H.~A.}\ \bibnamefont {Makse}},\ }\bibfield
  {title} {\bibinfo {title} {Edwards statistical mechanics for jammed granular
  matter},\ }\href {https://doi.org/10.1103/RevModPhys.90.015006} {\bibfield
  {journal} {\bibinfo  {journal} {Rev. Mod. Phys.}\ }\textbf {\bibinfo {volume}
  {90}},\ \bibinfo {pages} {015006} (\bibinfo {year} {2018})}\BibitemShut
  {NoStop}%
\bibitem [{\citenamefont {Papadopoulos}\ \emph {et~al.}(2016)\citenamefont
  {Papadopoulos}, \citenamefont {Puckett}, \citenamefont {Daniels},\ and\
  \citenamefont {Bassett}}]{Papadopoulos2016}%
  \BibitemOpen
  \bibfield  {author} {\bibinfo {author} {\bibfnamefont {L.}~\bibnamefont
  {Papadopoulos}}, \bibinfo {author} {\bibfnamefont {J.~G.}\ \bibnamefont
  {Puckett}}, \bibinfo {author} {\bibfnamefont {K.~E.}\ \bibnamefont
  {Daniels}},\ and\ \bibinfo {author} {\bibfnamefont {D.~S.}\ \bibnamefont
  {Bassett}},\ }\bibfield  {title} {\bibinfo {title} {Evolution of network
  architecture in a granular material under compression},\ }\href
  {https://doi.org/10.1103/PhysRevE.94.032908} {\bibfield  {journal} {\bibinfo
  {journal} {Phys. Rev. E}\ }\textbf {\bibinfo {volume} {94}},\ \bibinfo
  {pages} {032908} (\bibinfo {year} {2016})}\BibitemShut {NoStop}%
\bibitem [{\citenamefont {Krishnaraj}\ and\ \citenamefont
  {Nott}(2020)}]{Krishnaraj2020}%
  \BibitemOpen
  \bibfield  {author} {\bibinfo {author} {\bibfnamefont {K.~P.}\ \bibnamefont
  {Krishnaraj}}\ and\ \bibinfo {author} {\bibfnamefont {P.~R.}\ \bibnamefont
  {Nott}},\ }\bibfield  {title} {\bibinfo {title} {Coherent force chains in
  disordered granular materials emerge from a percolation of quasilinear
  clusters},\ }\href {https://doi.org/10.1103/PhysRevLett.124.198002}
  {\bibfield  {journal} {\bibinfo  {journal} {Phys. Rev. Lett.}\ }\textbf
  {\bibinfo {volume} {124}},\ \bibinfo {pages} {198002} (\bibinfo {year}
  {2020})}\BibitemShut {NoStop}%
\bibitem [{\citenamefont {Mandal}\ \emph {et~al.}(2022)\citenamefont {Mandal},
  \citenamefont {Casert},\ and\ \citenamefont {Sollich}}]{Mandal2022}%
  \BibitemOpen
  \bibfield  {author} {\bibinfo {author} {\bibfnamefont {R.}~\bibnamefont
  {Mandal}}, \bibinfo {author} {\bibfnamefont {C.}~\bibnamefont {Casert}},\
  and\ \bibinfo {author} {\bibfnamefont {P.}~\bibnamefont {Sollich}},\
  }\bibfield  {title} {\bibinfo {title} {Robust prediction of force chains in
  jammed solids using graph neural networks},\ }\href
  {https://doi.org/10.1038/s41467-022-31732-3} {\bibfield  {journal} {\bibinfo
  {journal} {Nat. Comm.}\ }\textbf {\bibinfo {volume} {13}},\ \bibinfo {pages}
  {4424} (\bibinfo {year} {2022})}\BibitemShut {NoStop}%
\bibitem [{\citenamefont {Alexander}(1998)}]{Alexander1998}%
  \BibitemOpen
  \bibfield  {author} {\bibinfo {author} {\bibfnamefont {S.}~\bibnamefont
  {Alexander}},\ }\bibfield  {title} {\bibinfo {title} {Amorphous solids: Their
  structure, lattice dynamics and elasticity},\ }\href
  {https://doi.org/10.1016/S0370-1573(97)00069-0} {\bibfield  {journal}
  {\bibinfo  {journal} {Phys. Rep.}\ }\textbf {\bibinfo {volume} {296}},\
  \bibinfo {pages} {65} (\bibinfo {year} {1998})}\BibitemShut {NoStop}%
\bibitem [{\citenamefont {Wyart}(2005)}]{Wyart2005}%
  \BibitemOpen
  \bibfield  {author} {\bibinfo {author} {\bibfnamefont {M.}~\bibnamefont
  {Wyart}},\ }\bibfield  {title} {\bibinfo {title} {On the rigidity of
  amorphous solids},\ }\href {https://doi.org/10.1051/anphys:2006003}
  {\bibfield  {journal} {\bibinfo  {journal} {Ann. Phys.}\ }\textbf {\bibinfo
  {volume} {30}},\ \bibinfo {pages} {1} (\bibinfo {year} {2005})}\BibitemShut
  {NoStop}%
\bibitem [{\citenamefont {Forsyth}\ \emph {et~al.}(2001)\citenamefont
  {Forsyth}, \citenamefont {Hutton}, \citenamefont {Osborne},\ and\
  \citenamefont {Rhodes}}]{Forsyth01}%
  \BibitemOpen
  \bibfield  {author} {\bibinfo {author} {\bibfnamefont {A.~J.}\ \bibnamefont
  {Forsyth}}, \bibinfo {author} {\bibfnamefont {S.~R.}\ \bibnamefont {Hutton}},
  \bibinfo {author} {\bibfnamefont {C.~F.}\ \bibnamefont {Osborne}},\ and\
  \bibinfo {author} {\bibfnamefont {M.~J.}\ \bibnamefont {Rhodes}},\ }\bibfield
   {title} {\bibinfo {title} {Effects of interparticle force on the packing of
  spherical granular material},\ }\href
  {https://doi.org/10.1103/PhysRevLett.87.244301} {\bibfield  {journal}
  {\bibinfo  {journal} {Phys. Rev. Lett.}\ }\textbf {\bibinfo {volume} {87}},\
  \bibinfo {pages} {244301} (\bibinfo {year} {2001})}\BibitemShut {NoStop}%
\bibitem [{\citenamefont {Gans}\ \emph {et~al.}(2024)\citenamefont {Gans},
  \citenamefont {Dalloz-Dubrujeaud}, \citenamefont {Nicolas},\ and\
  \citenamefont {Aussillous}}]{Gans24}%
  \BibitemOpen
  \bibfield  {author} {\bibinfo {author} {\bibfnamefont {A.}~\bibnamefont
  {Gans}}, \bibinfo {author} {\bibfnamefont {B.}~\bibnamefont
  {Dalloz-Dubrujeaud}}, \bibinfo {author} {\bibfnamefont {M.}~\bibnamefont
  {Nicolas}},\ and\ \bibinfo {author} {\bibfnamefont {P.}~\bibnamefont
  {Aussillous}},\ }\bibfield  {title} {\bibinfo {title} {Discharge flow of a
  cohesive granular media from a silo},\ }\href
  {https://doi.org/10.1103/PhysRevLett.133.238201} {\bibfield  {journal}
  {\bibinfo  {journal} {Phys. Rev. Lett.}\ }\textbf {\bibinfo {volume} {133}},\
  \bibinfo {pages} {238201} (\bibinfo {year} {2024})}\BibitemShut {NoStop}%
\bibitem [{\citenamefont {Fujio}\ \emph {et~al.}(2024)\citenamefont {Fujio},
  \citenamefont {Yokota}, \citenamefont {Tani},\ and\ \citenamefont
  {Kurita}}]{Fujio2024}%
  \BibitemOpen
  \bibfield  {author} {\bibinfo {author} {\bibfnamefont {H.}~\bibnamefont
  {Fujio}}, \bibinfo {author} {\bibfnamefont {H.}~\bibnamefont {Yokota}},
  \bibinfo {author} {\bibfnamefont {M.}~\bibnamefont {Tani}},\ and\ \bibinfo
  {author} {\bibfnamefont {R.}~\bibnamefont {Kurita}},\ }\bibfield  {title}
  {\bibinfo {title} {Gel-like mechanisms of durability and deformability in wet
  granular systems},\ }\href {https://doi.org/10.1038/s42005-023-01518-0}
  {\bibfield  {journal} {\bibinfo  {journal} {Nat. Comm.}\ }\textbf {\bibinfo
  {volume} {15}},\ \bibinfo {pages} {1234} (\bibinfo {year}
  {2024})}\BibitemShut {NoStop}%
\bibitem [{\citenamefont {Shrivastava}\ \emph {et~al.}(2025)\citenamefont
  {Shrivastava}, \citenamefont {Dayal},\ and\ \citenamefont
  {Noh}}]{Shrivastava25}%
  \BibitemOpen
  \bibfield  {author} {\bibinfo {author} {\bibfnamefont {A.}~\bibnamefont
  {Shrivastava}}, \bibinfo {author} {\bibfnamefont {K.}~\bibnamefont {Dayal}},\
  and\ \bibinfo {author} {\bibfnamefont {H.~Y.}\ \bibnamefont {Noh}},\
  }\bibfield  {title} {\bibinfo {title} {The roles of size, packing, and
  cohesion in the emergence of force chains in granular packings},\ }\href
  {https://doi.org/10.1115/1.4068059} {\bibfield  {journal} {\bibinfo
  {journal} {J. Appl. Mech.}\ }\textbf {\bibinfo {volume} {92}},\ \bibinfo
  {pages} {061003} (\bibinfo {year} {2025})}\BibitemShut {NoStop}%
\bibitem [{\citenamefont {Ness}\ and\ \citenamefont {Fielding}(2025)}]{Ness25}%
  \BibitemOpen
  \bibfield  {author} {\bibinfo {author} {\bibfnamefont {C.}~\bibnamefont
  {Ness}}\ and\ \bibinfo {author} {\bibfnamefont {S.~M.}\ \bibnamefont
  {Fielding}},\ }\bibfield  {title} {\bibinfo {title} {Nonmonotonic
  constitutive curves and shear banding in dry and wet granular flows},\ }\href
  {https://doi.org/10.1103/PhysRevLett.134.038201} {\bibfield  {journal}
  {\bibinfo  {journal} {Phys. Rev. Lett.}\ }\textbf {\bibinfo {volume} {134}},\
  \bibinfo {pages} {038201} (\bibinfo {year} {2025})}\BibitemShut {NoStop}%
\bibitem [{\citenamefont {Sharma}\ and\ \citenamefont
  {Sauret}(2025)}]{Sharma2025}%
  \BibitemOpen
  \bibfield  {author} {\bibinfo {author} {\bibfnamefont {R.}~\bibnamefont
  {Sharma}}\ and\ \bibinfo {author} {\bibfnamefont {A.}~\bibnamefont
  {Sauret}},\ }\bibfield  {title} {\bibinfo {title} {Experimental models for
  cohesive granular materials: a review},\ }\href
  {https://doi.org/10.1039/D4SM01324G} {\bibfield  {journal} {\bibinfo
  {journal} {Soft Matter}\ }\textbf {\bibinfo {volume} {21}},\ \bibinfo {pages}
  {450} (\bibinfo {year} {2025})}\BibitemShut {NoStop}%
\bibitem [{\citenamefont {Hornbaker}\ \emph {et~al.}(1997)\citenamefont
  {Hornbaker}, \citenamefont {Albert}, \citenamefont {Albert}, \citenamefont
  {Barabási},\ and\ \citenamefont {Schiffer}}]{Hornbaker1997}%
  \BibitemOpen
  \bibfield  {author} {\bibinfo {author} {\bibfnamefont {D.~J.}\ \bibnamefont
  {Hornbaker}}, \bibinfo {author} {\bibfnamefont {R.}~\bibnamefont {Albert}},
  \bibinfo {author} {\bibfnamefont {I.}~\bibnamefont {Albert}}, \bibinfo
  {author} {\bibfnamefont {A.-L.}\ \bibnamefont {Barabási}},\ and\ \bibinfo
  {author} {\bibfnamefont {P.}~\bibnamefont {Schiffer}},\ }\bibfield  {title}
  {\bibinfo {title} {What keeps sandcastles standing?},\ }\href
  {https://doi.org/10.1038/42831} {\bibfield  {journal} {\bibinfo  {journal}
  {Nature}\ }\textbf {\bibinfo {volume} {387}},\ \bibinfo {pages} {765}
  (\bibinfo {year} {1997})}\BibitemShut {NoStop}%
\bibitem [{\citenamefont {Selmani}\ \emph {et~al.}(2024)\citenamefont
  {Selmani}, \citenamefont {Besnard}, \citenamefont {Ould El~Moctar},
  \citenamefont {Dupont},\ and\ \citenamefont {Valance}}]{Selmani24}%
  \BibitemOpen
  \bibfield  {author} {\bibinfo {author} {\bibfnamefont {H.}~\bibnamefont
  {Selmani}}, \bibinfo {author} {\bibfnamefont {J.~B.}\ \bibnamefont
  {Besnard}}, \bibinfo {author} {\bibfnamefont {A.}~\bibnamefont {Ould
  El~Moctar}}, \bibinfo {author} {\bibfnamefont {P.}~\bibnamefont {Dupont}},\
  and\ \bibinfo {author} {\bibfnamefont {A.}~\bibnamefont {Valance}},\
  }\bibfield  {title} {\bibinfo {title} {Experimental study of particle impact
  on cohesive granular packing},\ }\href
  {https://doi.org/10.1103/PhysRevE.110.014901} {\bibfield  {journal} {\bibinfo
   {journal} {Phys. Rev. E}\ }\textbf {\bibinfo {volume} {110}},\ \bibinfo
  {pages} {014901} (\bibinfo {year} {2024})}\BibitemShut {NoStop}%
\bibitem [{\citenamefont {Palzer}(2005)}]{Palzer2005}%
  \BibitemOpen
  \bibfield  {author} {\bibinfo {author} {\bibfnamefont {S.}~\bibnamefont
  {Palzer}},\ }\bibfield  {title} {\bibinfo {title} {The effect of glass
  transition on the desired and undesired agglomeration of amorphous food
  powders},\ }\href {https://doi.org/10.1016/j.ces.2005.02.015} {\bibfield
  {journal} {\bibinfo  {journal} {Chem. Eng. Sci.}\ }\textbf {\bibinfo {volume}
  {60}},\ \bibinfo {pages} {3959} (\bibinfo {year} {2005})}\BibitemShut
  {NoStop}%
\bibitem [{\citenamefont {Hartmann}\ and\ \citenamefont
  {Palzer}(2011)}]{Hartmann2011}%
  \BibitemOpen
  \bibfield  {author} {\bibinfo {author} {\bibfnamefont {M.}~\bibnamefont
  {Hartmann}}\ and\ \bibinfo {author} {\bibfnamefont {S.}~\bibnamefont
  {Palzer}},\ }\bibfield  {title} {\bibinfo {title} {Caking of amorphous
  powders - material aspects, modelling and applications},\ }\href
  {https://doi.org/10.1016/j.powtec.2010.04.014} {\bibfield  {journal}
  {\bibinfo  {journal} {Powder Technol.}\ }\textbf {\bibinfo {volume} {206}},\
  \bibinfo {pages} {112} (\bibinfo {year} {2011})}\BibitemShut {NoStop}%
\bibitem [{\citenamefont {Zafar}\ \emph {et~al.}(2017)\citenamefont {Zafar},
  \citenamefont {Vivacqua}, \citenamefont {Calvert}, \citenamefont {Ghadiri},\
  and\ \citenamefont {Cleaver}}]{Zafar2017}%
  \BibitemOpen
  \bibfield  {author} {\bibinfo {author} {\bibfnamefont {U.}~\bibnamefont
  {Zafar}}, \bibinfo {author} {\bibfnamefont {V.}~\bibnamefont {Vivacqua}},
  \bibinfo {author} {\bibfnamefont {G.}~\bibnamefont {Calvert}}, \bibinfo
  {author} {\bibfnamefont {M.}~\bibnamefont {Ghadiri}},\ and\ \bibinfo {author}
  {\bibfnamefont {J.~A.~S.}\ \bibnamefont {Cleaver}},\ }\bibfield  {title}
  {\bibinfo {title} {A review of bulk powder caking},\ }\href
  {https://doi.org/10.1016/j.powtec.2017.02.024} {\bibfield  {journal}
  {\bibinfo  {journal} {Powder Technol.}\ }\textbf {\bibinfo {volume} {313}},\
  \bibinfo {pages} {389} (\bibinfo {year} {2017})}\BibitemShut {NoStop}%
\bibitem [{\citenamefont {Chen}\ \emph {et~al.}(2019)\citenamefont {Chen},
  \citenamefont {Zhang}, \citenamefont {Dong}, \citenamefont {Luo},
  \citenamefont {Kang}, \citenamefont {Li}, \citenamefont {Wang},\ and\
  \citenamefont {Gong}}]{Chen2019}%
  \BibitemOpen
  \bibfield  {author} {\bibinfo {author} {\bibfnamefont {M.}~\bibnamefont
  {Chen}}, \bibinfo {author} {\bibfnamefont {D.}~\bibnamefont {Zhang}},
  \bibinfo {author} {\bibfnamefont {W.}~\bibnamefont {Dong}}, \bibinfo {author}
  {\bibfnamefont {Z.}~\bibnamefont {Luo}}, \bibinfo {author} {\bibfnamefont
  {C.}~\bibnamefont {Kang}}, \bibinfo {author} {\bibfnamefont {H.}~\bibnamefont
  {Li}}, \bibinfo {author} {\bibfnamefont {G.}~\bibnamefont {Wang}},\ and\
  \bibinfo {author} {\bibfnamefont {J.}~\bibnamefont {Gong}},\ }\bibfield
  {title} {\bibinfo {title} {Amorphous and humidity caking: A review},\ }\href
  {https://doi.org/10.1016/j.cjche.2019.02.004} {\bibfield  {journal} {\bibinfo
   {journal} {Chin. J. Chem. Eng.}\ }\textbf {\bibinfo {volume} {27}},\
  \bibinfo {pages} {1429} (\bibinfo {year} {2019})}\BibitemShut {NoStop}%
\bibitem [{\citenamefont {Zaccone}(2022)}]{Zaccone2022}%
  \BibitemOpen
  \bibfield  {author} {\bibinfo {author} {\bibfnamefont {A.}~\bibnamefont
  {Zaccone}},\ }\bibfield  {title} {\bibinfo {title} {Explicit analytical
  solution for random close packing in $d=2$ and $d=3$},\ }\href
  {https://doi.org/10.1103/PhysRevLett.128.028002} {\bibfield  {journal}
  {\bibinfo  {journal} {Phys. Rev. Lett.}\ }\textbf {\bibinfo {volume} {128}},\
  \bibinfo {pages} {028002} (\bibinfo {year} {2022})}\BibitemShut {NoStop}%
\bibitem [{\citenamefont {Middleton}(2022)}]{Middleton2022}%
  \BibitemOpen
  \bibfield  {author} {\bibinfo {author} {\bibfnamefont {C.}~\bibnamefont
  {Middleton}},\ }\bibfield  {title} {\bibinfo {title} {Jamming connects
  granulation and flow},\ }\href {https://doi.org/10.1063/PT.3.5077} {\bibfield
   {journal} {\bibinfo  {journal} {Physics Today}\ }\textbf {\bibinfo {volume}
  {75}},\ \bibinfo {pages} {19} (\bibinfo {year} {2022})}\BibitemShut {NoStop}%
\bibitem [{\citenamefont {Xu}\ \emph {et~al.}(2025)\citenamefont {Xu},
  \citenamefont {Ma}, \citenamefont {Fu},\ and\ \citenamefont {Jiao}}]{Xu2025}%
  \BibitemOpen
  \bibfield  {author} {\bibinfo {author} {\bibfnamefont {W.}~\bibnamefont
  {Xu}}, \bibinfo {author} {\bibfnamefont {Z.}~\bibnamefont {Ma}}, \bibinfo
  {author} {\bibfnamefont {J.}~\bibnamefont {Fu}},\ and\ \bibinfo {author}
  {\bibfnamefont {Y.}~\bibnamefont {Jiao}},\ }\bibfield  {title} {\bibinfo
  {title} {Percolation-based mean field theory for disordered particle
  packings},\ }\href {https://doi.org/10.1016/j.powtec.2025.121088} {\bibfield
  {journal} {\bibinfo  {journal} {Powder Technol.}\ ,\ \bibinfo {pages}
  {121088}} (\bibinfo {year} {2025})}\BibitemShut {NoStop}%
\bibitem [{\citenamefont {Braz}\ \emph {et~al.}(2022)\citenamefont {Braz},
  \citenamefont {Matias}, \citenamefont {Forny}, \citenamefont {Pasche},
  \citenamefont {Meunier}, \citenamefont {Engmann},\ and\ \citenamefont
  {Araújo}}]{Braz2022}%
  \BibitemOpen
  \bibfield  {author} {\bibinfo {author} {\bibfnamefont {V.~C.}\ \bibnamefont
  {Braz}}, \bibinfo {author} {\bibfnamefont {A.~F.~V.}\ \bibnamefont {Matias}},
  \bibinfo {author} {\bibfnamefont {L.}~\bibnamefont {Forny}}, \bibinfo
  {author} {\bibfnamefont {D.}~\bibnamefont {Pasche}}, \bibinfo {author}
  {\bibfnamefont {V.}~\bibnamefont {Meunier}}, \bibinfo {author} {\bibfnamefont
  {J.}~\bibnamefont {Engmann}},\ and\ \bibinfo {author} {\bibfnamefont
  {N.~A.~M.}\ \bibnamefont {Araújo}},\ }\bibfield  {title} {\bibinfo {title}
  {Percolation-based simulation to predict caking kinetics of polydisperse
  amorphous powders},\ }\href {https://doi.org/10.1016/j.powtec.2022.117248}
  {\bibfield  {journal} {\bibinfo  {journal} {Powder Technol.}\ }\textbf
  {\bibinfo {volume} {400}},\ \bibinfo {pages} {117248} (\bibinfo {year}
  {2022})}\BibitemShut {NoStop}%
\bibitem [{\citenamefont {Simões}\ \emph {et~al.}(2022)\citenamefont
  {Simões}, \citenamefont {Matias}, \citenamefont {Braz}, \citenamefont
  {Pasche}, \citenamefont {Meunier}, \citenamefont {Engmann},\ and\
  \citenamefont {Araújo}}]{Simoes2022}%
  \BibitemOpen
  \bibfield  {author} {\bibinfo {author} {\bibfnamefont {T.~S. A.~N.}\
  \bibnamefont {Simões}}, \bibinfo {author} {\bibfnamefont {A.~F.~V.}\
  \bibnamefont {Matias}}, \bibinfo {author} {\bibfnamefont {V.~C.}\
  \bibnamefont {Braz}}, \bibinfo {author} {\bibfnamefont {D.}~\bibnamefont
  {Pasche}}, \bibinfo {author} {\bibfnamefont {V.}~\bibnamefont {Meunier}},
  \bibinfo {author} {\bibfnamefont {J.}~\bibnamefont {Engmann}},\ and\ \bibinfo
  {author} {\bibfnamefont {N.~A.~M.}\ \bibnamefont {Araújo}},\ }\bibfield
  {title} {\bibinfo {title} {Effect of temperature shocks on the caking of
  moisture-sensitive amorphous powders},\ }\href
  {https://doi.org/10.1016/j.powtec.2022.117799} {\bibfield  {journal}
  {\bibinfo  {journal} {Powder Technol.}\ }\textbf {\bibinfo {volume} {409}},\
  \bibinfo {pages} {117799} (\bibinfo {year} {2022})}\BibitemShut {NoStop}%
\bibitem [{\citenamefont {Artime}\ \emph {et~al.}(2024)\citenamefont {Artime},
  \citenamefont {Grassia}, \citenamefont {Domenico}, \citenamefont {Gleeson},
  \citenamefont {Makse}, \citenamefont {Mangioni}, \citenamefont {Perc},\ and\
  \citenamefont {Radicchi}}]{Artime2024}%
  \BibitemOpen
  \bibfield  {author} {\bibinfo {author} {\bibfnamefont {O.}~\bibnamefont
  {Artime}}, \bibinfo {author} {\bibfnamefont {M.}~\bibnamefont {Grassia}},
  \bibinfo {author} {\bibfnamefont {M.~D.}\ \bibnamefont {Domenico}}, \bibinfo
  {author} {\bibfnamefont {J.~P.}\ \bibnamefont {Gleeson}}, \bibinfo {author}
  {\bibfnamefont {H.~A.}\ \bibnamefont {Makse}}, \bibinfo {author}
  {\bibfnamefont {G.}~\bibnamefont {Mangioni}}, \bibinfo {author}
  {\bibfnamefont {M.}~\bibnamefont {Perc}},\ and\ \bibinfo {author}
  {\bibfnamefont {F.}~\bibnamefont {Radicchi}},\ }\bibfield  {title} {\bibinfo
  {title} {Robustness and resilience of complex networks},\ }\href
  {https://doi.org/10.1038/s42254-023-00676-y} {\bibfield  {journal} {\bibinfo
  {journal} {Nat. Rev. Phys.}\ }\textbf {\bibinfo {volume} {6}},\ \bibinfo
  {pages} {114} (\bibinfo {year} {2024})}\BibitemShut {NoStop}%
\bibitem [{\citenamefont {Araújo}\ \emph {et~al.}(2014)\citenamefont
  {Araújo}, \citenamefont {Grassberger}, \citenamefont {Kahng}, \citenamefont
  {Schrenk},\ and\ \citenamefont {Ziff}}]{Araujo2014b}%
  \BibitemOpen
  \bibfield  {author} {\bibinfo {author} {\bibfnamefont {N.~A.~M.}\
  \bibnamefont {Araújo}}, \bibinfo {author} {\bibfnamefont {P.}~\bibnamefont
  {Grassberger}}, \bibinfo {author} {\bibfnamefont {B.}~\bibnamefont {Kahng}},
  \bibinfo {author} {\bibfnamefont {K.~J.}\ \bibnamefont {Schrenk}},\ and\
  \bibinfo {author} {\bibfnamefont {R.~M.}\ \bibnamefont {Ziff}},\ }\bibfield
  {title} {\bibinfo {title} {Recent advances and open challenges in
  percolation},\ }\href {https://doi.org/10.1140/epjst/e2014-02266-y}
  {\bibfield  {journal} {\bibinfo  {journal} {Eur. Phys. J. Spec. Top.}\
  }\textbf {\bibinfo {volume} {223}},\ \bibinfo {pages} {2307} (\bibinfo {year}
  {2014})}\BibitemShut {NoStop}%
\bibitem [{\citenamefont {Wood}\ and\ \citenamefont
  {Maeda}(2008)}]{MuirWood2008}%
  \BibitemOpen
  \bibfield  {author} {\bibinfo {author} {\bibfnamefont {D.~M.}\ \bibnamefont
  {Wood}}\ and\ \bibinfo {author} {\bibfnamefont {K.}~\bibnamefont {Maeda}},\
  }\bibfield  {title} {\bibinfo {title} {Changing grading of soil: effect on
  critical states},\ }\href {https://doi.org/10.1007/s11440-007-0041-0}
  {\bibfield  {journal} {\bibinfo  {journal} {Acta Geotechnica}\ }\textbf
  {\bibinfo {volume} {3}},\ \bibinfo {pages} {3} (\bibinfo {year}
  {2008})}\BibitemShut {NoStop}%
\bibitem [{\citenamefont {Az\'ema}\ \emph {et~al.}(2017)\citenamefont
  {Az\'ema}, \citenamefont {Linero}, \citenamefont {Estrada},\ and\
  \citenamefont {Lizcano}}]{Azema2017}%
  \BibitemOpen
  \bibfield  {author} {\bibinfo {author} {\bibfnamefont {E.}~\bibnamefont
  {Az\'ema}}, \bibinfo {author} {\bibfnamefont {S.}~\bibnamefont {Linero}},
  \bibinfo {author} {\bibfnamefont {N.}~\bibnamefont {Estrada}},\ and\ \bibinfo
  {author} {\bibfnamefont {A.}~\bibnamefont {Lizcano}},\ }\bibfield  {title}
  {\bibinfo {title} {Shear strength and microstructure of polydisperse
  packings: The effect of size span and shape of particle size distribution},\
  }\href {https://doi.org/10.1103/PhysRevE.96.022902} {\bibfield  {journal}
  {\bibinfo  {journal} {Phys. Rev. E}\ }\textbf {\bibinfo {volume} {96}},\
  \bibinfo {pages} {022902} (\bibinfo {year} {2017})}\BibitemShut {NoStop}%
\bibitem [{SM()}]{SM}%
  \BibitemOpen
  \href {...} {}\bibinfo {note} {See Supplemental Material at .}\BibitemShut
  {Stop}%
\bibitem [{\citenamefont {Molloy}\ and\ \citenamefont
  {Reed}(1995)}]{Molloy1995}%
  \BibitemOpen
  \bibfield  {author} {\bibinfo {author} {\bibfnamefont {M.}~\bibnamefont
  {Molloy}}\ and\ \bibinfo {author} {\bibfnamefont {B.}~\bibnamefont {Reed}},\
  }\bibfield  {title} {\bibinfo {title} {A critical point for random graphs
  with a given degree sequence},\ }\href
  {https://doi.org/10.1002/rsa.3240060204} {\bibfield  {journal} {\bibinfo
  {journal} {Random Struct. Algorithms}\ }\textbf {\bibinfo {volume} {6}},\
  \bibinfo {pages} {161} (\bibinfo {year} {1995})}\BibitemShut {NoStop}%
\bibitem [{\citenamefont {Srivastava}\ \emph {et~al.}(2012)\citenamefont
  {Srivastava}, \citenamefont {Mitra}, \citenamefont {Ganguly},\ and\
  \citenamefont {Peruani}}]{Peruani2012}%
  \BibitemOpen
  \bibfield  {author} {\bibinfo {author} {\bibfnamefont {A.}~\bibnamefont
  {Srivastava}}, \bibinfo {author} {\bibfnamefont {B.}~\bibnamefont {Mitra}},
  \bibinfo {author} {\bibfnamefont {N.}~\bibnamefont {Ganguly}},\ and\ \bibinfo
  {author} {\bibfnamefont {F.}~\bibnamefont {Peruani}},\ }\bibfield  {title}
  {\bibinfo {title} {Correlations in complex networks under attack},\ }\href
  {https://doi.org/10.1103/PhysRevE.86.036106} {\bibfield  {journal} {\bibinfo
  {journal} {Phys. Rev. E}\ }\textbf {\bibinfo {volume} {86}},\ \bibinfo
  {pages} {036106} (\bibinfo {year} {2012})}\BibitemShut {NoStop}%
\bibitem [{\citenamefont {Dorogovtsev}\ and\ \citenamefont
  {Mendes}(2022)}]{Dorogovtsev2022}%
  \BibitemOpen
  \bibfield  {author} {\bibinfo {author} {\bibfnamefont {S.~N.}\ \bibnamefont
  {Dorogovtsev}}\ and\ \bibinfo {author} {\bibfnamefont {J.~F.~F.}\
  \bibnamefont {Mendes}},\ }\href
  {https://doi.org/10.1093/oso/9780199695119.001.0001} {\emph {\bibinfo {title}
  {The Nature of Complex Networks}}}\ (\bibinfo  {publisher} {Oxford University
  Press},\ \bibinfo {year} {2022})\BibitemShut {NoStop}%
\bibitem [{\citenamefont {Newman}(2010)}]{Newman2010}%
  \BibitemOpen
  \bibfield  {author} {\bibinfo {author} {\bibfnamefont {M.}~\bibnamefont
  {Newman}},\ }\href
  {https://doi.org/10.1093/acprof:oso/9780199206650.001.0001} {\emph {\bibinfo
  {title} {Networks: An Introduction}}}\ (\bibinfo  {publisher} {Oxford
  University Press},\ \bibinfo {year} {2010})\BibitemShut {NoStop}%
\bibitem [{\citenamefont {Barthélemy}(2011)}]{Barthelemy2011}%
  \BibitemOpen
  \bibfield  {author} {\bibinfo {author} {\bibfnamefont {M.}~\bibnamefont
  {Barthélemy}},\ }\bibfield  {title} {\bibinfo {title} {Spatial networks},\
  }\href {https://doi.org/10.1016/j.physrep.2010.11.002} {\bibfield  {journal}
  {\bibinfo  {journal} {Phys. Rep.}\ }\textbf {\bibinfo {volume} {499}},\
  \bibinfo {pages} {1} (\bibinfo {year} {2011})}\BibitemShut {NoStop}%
\end{thebibliography}%
\end{document}